\documentclass[aps,twocolumn,showkeys,superscriptaddress,amsmath,amssymb]{revtex4}

\usepackage{graphicx}
\usepackage{dcolumn}
\usepackage{bm}

\begin{document}

\title{Mechanism for the decomposition of lithium borohydride}

\author{Khang Hoang}
\affiliation{Materials Department, University of California, Santa Barbara, California 93106, USA}%
\affiliation{Computational Materials Science Center, George Mason University, Fairfax, Virginia 22030, USA}%
\author{Chris G. Van de Walle}
\affiliation{Materials Department, University of California, Santa Barbara, California 93106, USA}%


\begin{abstract}

We report first-principles density functional theory studies of native defects in lithium borohydride (LiBH$_{4}$), a potential material for hydrogen storage. Based on our detailed analysis of the structure, energetics, and migration of lithium-, boron-, and hydrogen-related defects, we propose a specific mechanism for the decomposition and dehydrogenation of LiBH$_{4}$ that involves mass transport mediated by native defects. In this mechanism, LiBH$_{4}$ releases borane (BH$_{3}$) at the surface or interface, leaving the negatively charged hydrogen interstitial (H$_{i}^{-}$) in the material, which then acts as the nucleation site for LiH formation. The diffusion of H$_{i}^{-}$ in the bulk LiBH$_{4}$ is the rate-limiting step in the decomposition kinetics. Lithium vacancies and interstitials have low formation energies and are highly mobile. These defects are responsible for maintaining local charge neutrality as other charged defects migrating along the material, and assisting in the formation of LiH. In light of this mechanism, we discuss the effects of metal additives on hydrogen desorption kinetics.

\end{abstract}


\maketitle

\section{\label{sec:intro}Introduction}

Lithium borohydride (LiBH$_{4}$) is a potential candidate for hydrogen storage because of its high hydrogen density (18.4 wt\%) \cite{Zuttel20031}. The major drawbacks of LiBH$_{4}$ that prevent it from practical use are its high decomposition temperature (370$^\circ$C) \cite{mauron2008} and slow hydrogen desorption kinetics. Although it has been reported that incorporation of some metal additives into the material can lower the decomposition temperature and enhance the kinetics \cite{Zuttel20031,en4010185,li2011}, the mechanism behind the decomposition and dehydrogenation processes and the role of the metal additives are not really understood. As demonstrated in our previous work for other complex hydrides, first-principles calculations based on density functional theory can produce such an understanding by providing insights on the atomistic mechanism involved in mass transport and hydrogen release in the material \cite{peles_prb_2007,hoang_prb_2009,wilson-short_prb_2009,hoang_angew}.

LiBH$_{4}$ undergoes a polymorphic transformation from the ordered low-temperature orthorhombic phase to the disordered high-temperature hexagonal phase around 104$^\circ$C \cite{Soulie2002}, and exhibits a slight hydrogen desorption of 0.3 wt\% between 100 and 200$^\circ$C \cite{Zuttel20031}. The melting occurs around 270$^\circ$C. The first and second significant hydrogen desorption peaks start at 320 and 400$^\circ$C, respectively, and the desorption reaches its maximum around 500$^\circ$C \cite{Zuttel20031}. The overall reaction for LiBH$_{4}$ decomposition can be expressed as the following equation:
\begin{equation}\label{eq;decomp}
\mathrm{Li}\mathrm{B}\mathrm{H}_{4} \rightarrow \mathrm{LiH} + \mathrm{B} + \frac{3}{2}\mathrm{H}_{2},
\end{equation}
which releases 13.8 wt\% hydrogen when LiBH$_{4}$ is heated up to 900$^\circ$C \cite{Zuttel20031}. The decomposition of LiBH$_{4}$ may, however, involve several intermediate steps. It has been reported that hydrogen desorption in LiBH$_{4}$ is accompanied by the release of gaseous diborane (B$_{2}$H$_{6}$) \cite{jp073783k}, which subsequently decomposes into B and H$_{2}$ at high temperatures \cite{cm100536a}. Some diboranes or higher boranes may, however, react with the not-yet-decomposed LiBH$_{4}$ to form Li$_{2}$B$_{12}$H$_{12}$ and possibly Li$_{2}$B$_{10}$H$_{10}$, which have been detected in experiments \cite{cm100536a,orimo:021920,jp710894t}.

Experimental data suggested that the decomposition and dehydrogenation of LiBH$_{4}$ involve hydrogen and/or boron mass transport mediated by native defects. However, there is no consensus among these experimental reports about the diffusing species involved in the decomposition process, which could be single H atoms, (BH$_{4}$)$^{-}$ units, or BH$_{3}$ units \cite{Borgschulte2008,shane2009,gremaud2009,borg_2010}. Regarding the activation energy for decomposition, Pendolino {\it et al.}~\cite{pendolino2009} reported a value of 1.22 eV for pure LiBH$_{4}$. For comparison, Z\"{u}ttel {\it et al.}~\cite{Zuttel20031} reported an activation energy of 1.62 eV for the decomposition of LiBH$_{4}$ mixed with SiO$_{2}$; other authors reported values of 1.36 and 1.06 eV for the ball-milled (3LiBH$_{4}$+MnCl$_{2}$) mixture \cite{Choudhury2009,varinLiBH4}.

In this paper we report first-principles studies of native point defects and defect complexes in LiBH$_{4}$. Based on our detailed analysis of the structure, energetics, and migration of the defects, we propose a specific atomistic mechanism that explains the decomposition and dehydrogenation of LiBH$_{4}$, and the effects of metal additives on hydrogen desorption kinetics. Comparison with the experimental work will be made throughout. Some results for hydrogen-related defects were reported previously \cite{hoang_prb_2009}, but are included here for completeness.

\section{\label{sec:metho}Methodology}

Our calculations are based on density functional theory, using the generalized-gradient approximation (GGA) \cite{GGA} and the projector-augmented wave method \cite{PAW1,PAW2} as implemented in the VASP code \cite{VASP1,VASP2,VASP3}. For defect calculations in LiBH$_{4}$ (orthorhombic $Pnma$, 24 atoms/unit cell) \cite{Soulie2002}, we used a (2$\times$2$\times$2) supercell which contains 192 atoms, and a 2$\times$2$\times$2 Monkhorst-Pack $\mathbf{k}$-point mesh \cite{monkhorst-pack}. The plane-wave basis-set cutoff was set to 400 eV and convergence with respect to self-consistent iterations was assumed when the total energy difference between cycles was less than 10$^{-4}$ eV and the residual forces were less than 0.01 eV/{\AA}. The migration of selected point defects in LiBH$_{4}$ was studied using the climbing-image nudged elastic band method (NEB) ~\cite{ci-neb}.

Throughout the paper we use defect formation energies to characterize different native defects in LiBH$_{4}$. The formation energy ($E^{f}$) of a defect is a crucial factor in determining its concentration. In thermal equilibrium, the concentration of the defect X at temperature $T$ can be obtained via the relation \cite{walle:3851,janotti2009}
\begin{equation}\label{eq;concen}
c(\mathrm{X})=N_{\mathrm{sites}}N_{\mathrm{config}}\mathrm{exp}[-E^{f}(\mathrm{X})/k_BT],
\end{equation}
where $N_{\mathrm{sites}}$ is the number of high-symmetry sites in the lattice per unit volume on which the defect can be incorporated, and $N_{\mathrm{config}}$ is the number of equivalent configurations per site. Note that the energy in Eq.~(\ref{eq;concen}) is, in principle, a free energy; however, the entropy and volume terms are often neglected because they are negligible at relevant experimental conditions \cite{janotti2009}. It is evident from Eq.~(\ref{eq;concen}) that defects with low formation energies will easily form and occur in high concentrations.

The formation energy of a defect X in charge state $q$ is defined as \cite{walle:3851}
\begin{eqnarray}\label{eq;eform}
\nonumber
E^f({\mathrm{X}}^q)=E_{\mathrm{tot}}({\mathrm{X}}^q)&-&E_{\mathrm{tot}}({\mathrm{bulk}})-\sum_{i}{n_i\mu_i} \\
&+&q(E_{\mathrm{V}}+\mu_{e}),
\end{eqnarray}
where $E_{\mathrm{tot}}(\mathrm{X}^{q})$ and $E_{\mathrm{tot}}(\mathrm{bulk})$ are, respectively, the total energies of a supercell containing the defect X, and of a supercell of the perfect bulk material; $\mu_{i}$ is the chemical potential of species $i$ (and is referenced to the standard state), and $n_{i}$ denotes the number of atoms of species $i$ that have been added ($n_{i}$$>$0) or removed ($n_{i}$$<$0) to form the defect. $\mu_{e}$ is the electron chemical potential, i.e., the Fermi level, referenced to the valence-band maximum in the bulk ($E_{\mathrm{V}}$).

The atomic chemical potentials $\mu_{i}$ are variables and can be chosen to represent experimental conditions. For defect calculations in LiBH$_{4}$, based on Eq.~(\ref{eq;decomp}) one can assume equilibrium of LiBH$_{4}$ with LiH and H$_{2}$, and the chemical potentials of Li, B, and H can be obtained from the equations that express the stability of LiH, H$_{2}$, and LiBH$_{4}$, which gives rise to $\mu_{\mathrm{Li}}$=$-$0.825, $\mu_{\mathrm{B}}$=$-$1.183, and $\mu_{\mathrm{H}}$=0 eV. This condition, however, corresponds to assuming equilibrium with LiH and H$_{2}$ at 0 K and 0 bar, and thus does not reflect the actual experimental conditions \cite{Zuttel20031}. One can also assume equilibrium with B and H$_{2}$, but this scenario is unlikely to occur during the decomposition of LiBH$_{4}$, given that both B and H$_{2}$ may not be formed directly from LiBH$_{4}$ but through the decomposition of the intermediates such as B$_{2}$H$_{6}$ \cite{cm100536a}. In the following presentation of defect formation energies, we assume equilibrium with LiH and B, which gives rise to $\mu_{\mathrm{Li}}$=$-$0.431, $\mu_{\mathrm{B}}$=0, and $\mu_{\mathrm{H}}$=$-$0.394 eV. This condition corresponds to assuming equilibrium with LiH and H$_{2}$ gas at 610 K and 1 bar \cite{H2gas}, which is close to the decomposition temperature (643 K) of LiBH$_{4}$ \cite{mauron2008}.

\section{\label{sec:defects}Point Defects and Complexes}

The compound can be regarded as an ordered arrangement of (Li)$^{+}$ and (BH$_{4}$)$^{-}$ units. In its electronic structure, the valence-band maximum (VBM) consists of the bonding states of B $p$ and H $s$, whereas the conduction-band minimum (CBM) consists predominantly of the antibonding states of B $p$ and H $s$. The calculated band gap is 7.01 eV \cite{hoang_prb_2009}. In such an insulating, large band gap material, native point defects are expected to exist in charged states other than neutral, and charge neutrality requires that defects with opposite charge states coexist in equal concentrations \cite{peles_prb_2007,hoang_prb_2009,wilson-short_prb_2009,hoang_angew}. We therefore investigated native defects in LiBH$_{4}$ in all possible charge states. Defect complexes are also considered, with special attention devoted to Frenkel pairs, i.e., interstitial-vacancy pairs of the same species. In the following, we analyze the structure, energetics, and migration of the defects in detail. The role of these defects in ionic conduction and the decomposition of LiBH$_{4}$ will be discussed in Sec.~\ref{sec:disc}.

\subsection{Hydrogen-related defects}

\begin{figure}
\begin{center}
\includegraphics[width=3.2in]{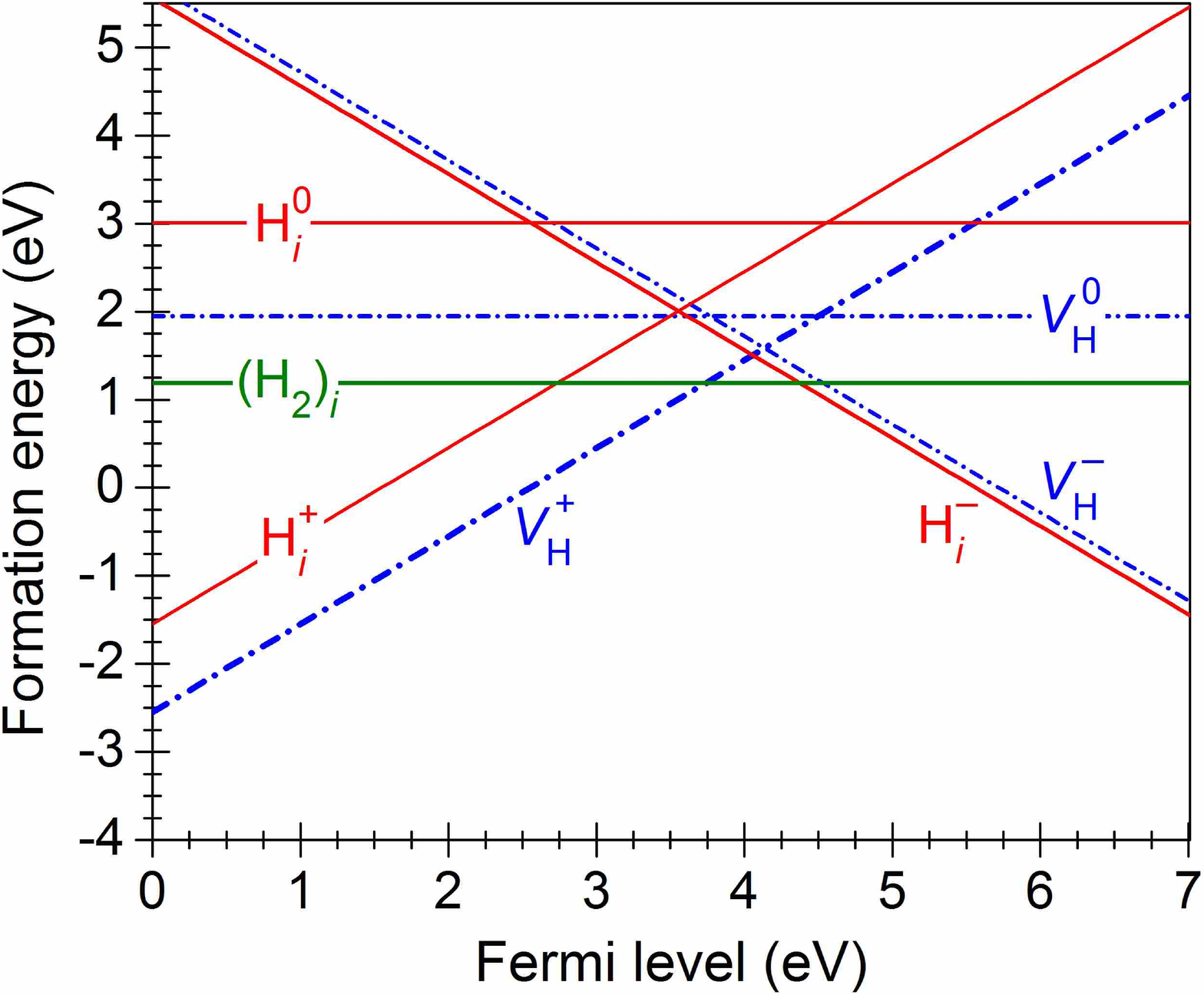}
\end{center}
\vspace{-0.2in}
\caption{(Color online) Calculated formation energies of hydrogen-related defects in LiBH$_{4}$, plotted as a function of Fermi energy with respect to the valence-band maximum.}\label{FE;H}
\end{figure}

\begin{figure}
\begin{center}
\includegraphics[width=3.2in]{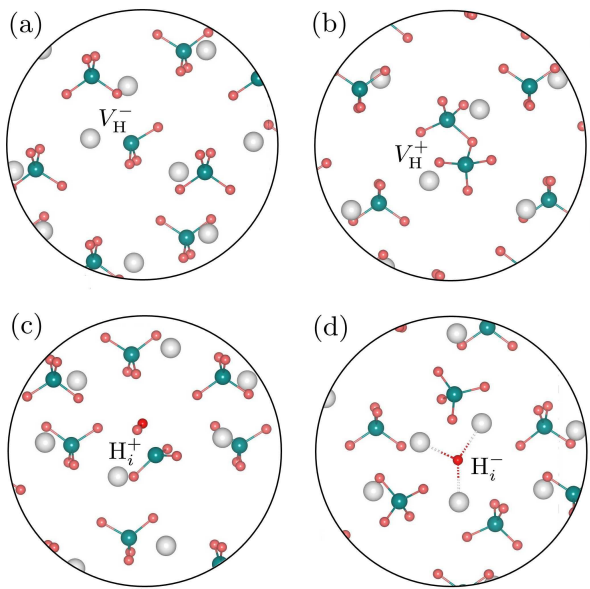}
\end{center}
\vspace{-0.15in}
\caption{(Color online) Structure of (a) $V_{\mathrm{H}}^{-}$, (b) $V_{\mathrm{H}}^{+}$, (c) H$_{i}^{+}$, and (d) H$_{i}^{-}$ in LiBH$_{4}$. Large (gray) spheres are Li, medium (blue) spheres B, and small (red) spheres H.}\label{struct}
\end{figure}

Figure \ref{FE;H} shows the calculated formation energies of hydrogen vacancies ($V_{\mathrm{H}}$), interstitials (H$_{i}$), and interstitial molecule (H$_{2}$)$_{i}$ in LiBH$_{4}$. We find that the positively charged hydrogen vacancy ($V_{\mathrm{H}}^{+}$) and negatively charged hydrogen interstitial (H$_{i}^{-}$) have the lowest formation energies over a wide range of Fermi-level values. The neutral hydrogen vacancy ($V_{\mathrm{H}}^{0}$) and interstitial (H$_{i}^{0}$) are energetically less favorable than their respective charged defects, which is a characteristic of negative-$U$ centers \cite{negativeU}. Near $\mu_{e}$=4.06 eV, where the formation energies of $V_{\mathrm{H}}^{+}$ and H$_{i}^{-}$ are equal, (H$_{2}$)$_{i}$ actually has the lowest formation energy.  The creation of $V_{\mathrm{H}}^{-}$ in LiBH$_{4}$ involves removing a proton (H$^{+}$) from the LiBH$_{4}$ supercell, and this results in a BH$_{3}$ unit; see Fig.~\ref{struct}(a). $V_{\mathrm{H}}^{+}$ is created by removing one H atom and an extra electron from the system, and this leads to formation of a BH$_{3}$-H-BH$_{3}$ complex (two BH$_{4}$ units sharing a common H atom); see Fig.~\ref{struct}(b). The addition of a proton to create H$_{i}^{+}$ results in a BH$_{5}$ complex; see Fig.~\ref{struct}(c). The creation of H$_{i}^{-}$, on the other hand, involves adding a H atom and an extra electron to the system, and the resulting H$_{i}^{-}$ stands next to three Li atoms with the average Li-H distance being 1.79 {\AA}; see Fig.~\ref{struct}(d). Finally, (H$_{2}$)$_{i}$ involves adding an H$_{2}$ molecule to the system. This interstitial molecule prefers to stay in the interstitial void with the calculated H$-$H bond length of 0.75 {\AA}, being equal to that calculated for an isolated H$_{2}$ molecule.

For the migration of H$_{i}^{+}$, H$_{i}^{-}$, $V_{\rm{H}}^{+}$, and $V_{\rm{H}}^{-}$, we find energy barriers of 0.65, 0.41, 0.91, and 1.32 eV, respectively. The energy barriers for H$_{i}^{+}$, $V_{\rm{H}}^{+}$, and $V_{\rm{H}}^{-}$ are relatively high because the diffusion of these defects involves breaking B$-$H bonds. For example, the diffusion of $V_{\rm{H}}^{-}$ involves moving an H atom from a BH$_{4}$ unit to the vacancy. The saddle-point configuration in this case consists of a H atom located midway between two BH$_{4}$ units (i.e., BH$_{4}$-H-BH$_{4}$), an energetically favorable situation for $V_{\rm{H}}^{+}$ but unfavorable for $V_{\rm{H}}^{-}$. H$_{i}^{-}$, on the other hand, loosely bonds to three Li atoms and therefore can diffuse more easily. The barrier for H$_{i}^{-}$ given here is slightly lower than the preliminary value reported previously \cite{hoang_prb_2009}.

Since hydrogen vacancies and interstitials are stable as oppositely charged defects, charge and mass conservation conditions suggest that these native defects may be created in the interior of the material in form of Frenkel pairs. Therefore, we have investigated the formation of hydrogen Frenkel pairs (H$_{i}^{+}$,$V_{\mathrm{H}}^{-}$) and (H$_{i}^{-}$,$V_{\mathrm{H}}^{+}$). We find that in these complexes the configurations of the individual defects are preserved. (H$_{i}^{+}$,$V_{\mathrm{H}}^{-}$) has a formation energy of 3.89 eV, and a binding energy of 0.28 eV with respect to the isolated constituents. The distance between the two defects in the pair (as measured by the B$-$B distance) is 3.55 {\AA}, compared to the B$-$B distance of 3.64 {\AA} in the bulk. (H$_{i}^{-}$,$V_{\mathrm{H}}^{+}$), on the other hand, has a formation energy of 2.28 eV and a binding energy of 0.73 eV. The distance from H$_{i}^{-}$ to the H atom near the center of $V_{\mathrm{H}}^{+}$ [{\it cf.} Fig.~\ref{struct}(b)] is 4.02 {\AA}. Thus, (H$_{i}^{-}$,$V_{\mathrm{H}}^{+}$) has a much lower formation energy than (H$_{i}^{+}$,$V_{\mathrm{H}}^{-}$), which is consistent with the results presented in Fig.~\ref{FE;H} where H$_{i}^{-}$ and $V_{\mathrm{H}}^{+}$ both have lower formation energies.

Hao and Sholl \cite{hao2009} recently reported first-principles calculations for hydrogen-related defects in LiBH$_{4}$, using a methodology similar to ours. The formation energies (evaluated with $\mu_{\mathrm{H}}$=0 eV) of the defects, except H$_{i}^{-}$, are in close agreement with our results obtained under condition (1) as reported in Table \ref{tab:libh} (to within 0.1 eV). For H$_{i}^{-}$, our calculated formation energy is lower by 0.3$-$0.4 eV, suggesting that the H$_{i}^{-}$ configuration we identified is more stable. Hao and Sholl also considered neutral hydrogen divacancies in LiBH$_{4}$ (denoted as $V_{\rm 2H}$) by removing two H atoms from the system, resulting in a B$_{2}$H$_{6}$ unit with an ethane-like geometry. The formation energy of $V_{\rm 2H}$ was reported to be 1.14 eV \cite{hao2009}, comparable to that (1.11 eV) obtained under the condition with $\mu_{\mathrm{H}}$=0 eV in our calculations. This divacancy can be regarded as a complex of $V_{\rm H}^{+}$ and $V_{\rm H}^{-}$ with a binding energy of 2.84 eV with respect to its individual constituents, although the structures of the individual defects are not preserved in the divacancy. Hao and Sholl \cite{hao2009} also noted that the hydrogen divacancy and interstitial hydrogen molecule have formation energies much lower than those of $V_{\rm H}^{+}$ and H$_{i}^{-}$.  The sum of their calculated formation energies for $V_{\rm 2H}$ and (H$_{2}$)$_{i}$ is 1.56 eV, close to the value of 1.51 eV in our calculations.  We have also explicitly calculated a $V_{\rm 2H}$-(H$_{2}$)$_{i}$ complex, finding a formation energy of 1.49 eV.  Note that the formation energy of this complex is independent of the chemical potentials. The $V_{\rm 2H}$-(H$_{2}$)$_{i}$ complex therefore has almost zero binding energy (0.02 eV) with respect to its constituents, suggesting that, once created, it would readily dissociate into $V_{\rm 2H}$ and (H$_{2}$)$_{i}$.

\subsection{Lithium-related defects}

\begin{figure}
\begin{center}
\includegraphics[width=3.2in]{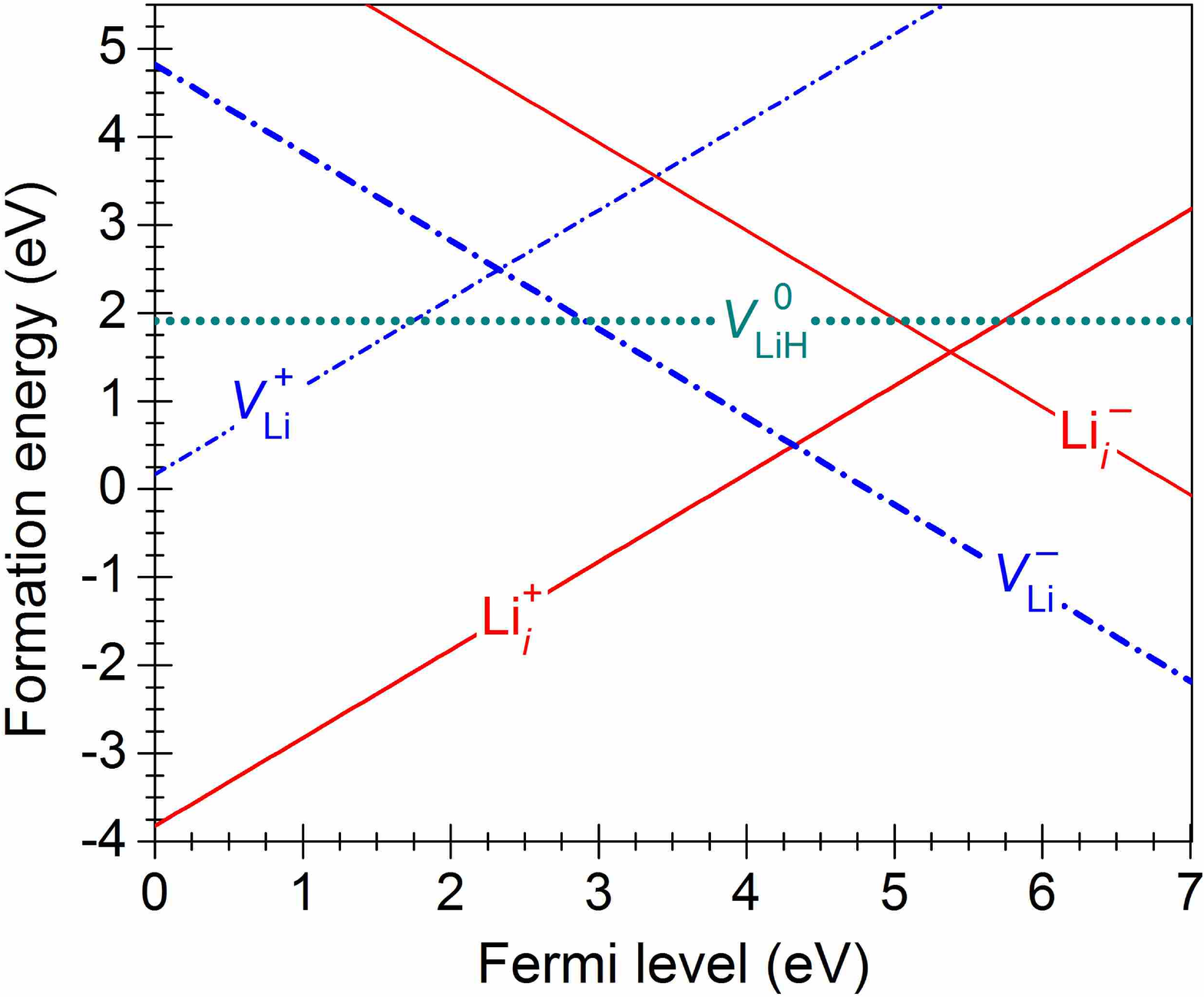}
\end{center}
\vspace{-0.2in}
\caption{(Color online) Calculated formation energies of lithium-related defects in LiBH$_{4}$, plotted as a function of Fermi energy with respect to the valence-band maximum.}\label{LiBH4;FE;Li}
\end{figure}

Figure \ref{LiBH4;FE;Li} shows the calculated formation energies of lithium vacancies ($V_{\mathrm{Li}}$), lithium interstitials (Li$_{i}$), and $V_{\mathrm{LiH}}$ (removing Li and H) in LiBH$_{4}$. Among these defects, we find that Li$_{i}^{+}$ and $V_{\mathrm{Li}}^{-}$ have the lowest formation energies for the entire range of Fermi-level values. These two defects have equal formation energies at $\mu_{e}$=4.32 eV. The creation of $V_{\mathrm{Li}}^{-}$ involves removing a Li$^{+}$ ion from the system. This causes very small changes to the lattice geometry near the void created by the removed ion. On the contrary, $V_{\mathrm{Li}}^{+}$, created by removing a Li atom and an extra electron, strongly disturbs the system. Besides the void, there are two BH$_{4}$ units that come close and form a B$_{2}$H$_{8}$ complex which can be identified as H$_{i}^{+}$ plus $V_{\mathrm{H}}^{+}$. $V_{\mathrm{Li}}^{+}$ thus can be regarded as a complex of $V_{\mathrm{Li}}^{-}$, H$_{i}^{+}$, and $V_{\mathrm{H}}^{+}$, with a binding energy of 0.56 eV with respect to its isolated constituents. Regarding the interstitials, Li$_{i}^{+}$ is created by adding a Li$^{+}$ ion to the system. Like $V_{\mathrm{Li}}^{-}$, Li$_{i}^{+}$ does not cause much disturbance to the lattice geometry of LiBH$_{4}$. Li$_{i}^{-}$, however, strongly disturbs the system by breaking B$-$H bonds and forming H and BH$_{3}$ units which can be identified as H$_{i}^{-}$ and $V_{\mathrm{H}}^{-}$, respectively. This defect can, therefore, be considered as a complex of Li$_{i}^{+}$, $V_{\mathrm{H}}^{-}$, and H$_{i}^{-}$, with a binding energy of 0.52 eV. Finally, $V_{\mathrm{LiH}}^{0}$ can be regarded as a complex of $V_{\mathrm{Li}}^{-}$ and $V_{\mathrm{H}}^{+}$.

The migration of Li$_{i}^{+}$ involves an energy barrier as low as 0.30 eV, and the migration of $V_{\mathrm{Li}}^{-}$ involves a barrier of 0.29 eV. These values are small, suggesting that these two defects are highly mobile. For Li$_{i}^{-}$, which can be considered as a complex of Li$_{i}^{+}$, $V_{\mathrm{H}}^{-}$, and H$_{i}^{-}$, the migration barrier is given by the least mobile constituent \cite{wilson-short_prb_2009}, i.e., 1.32 eV, the value for $V_{\mathrm{H}}^{-}$. Similarly, the estimated migration barrier of $V_{\mathrm{Li}}^{+}$ and $V_{\mathrm{LiH}}^{0}$ is 0.91 eV, the value for $V_{\mathrm{H}}^{+}$.

We also investigated possible formation of lithium Frenkel pairs. Since Li$_{i}^{-}$ and $V_{\mathrm{Li}}^{+}$ are complex defects, only (Li$_{i}^{+}$,$V_{\mathrm{Li}}^{-}$) was considered. This pair is, however, unstable at short pair distances because there is no energy barrier between Li$_{i}^{+}$ and the neighboring $V_{\mathrm{Li}}^{+}$. In order to avoid recombination, Li$_{i}^{+}$ and $V_{\mathrm{Li}}^{-}$ should be about 4.20 {\AA} away from each other.  At this distance, the binding energy of (Li$_{i}^{+}$,$V_{\mathrm{Li}}^{-}$) with respect to its isolated constituents is 0.04 eV (almost zero), and the formation energy is 0.95 eV. This indicates that once created (Li$_{i}^{+}$,$V_{\mathrm{Li}}^{-}$) will readily recombine or dissociate into Li$_{i}^{+}$ and $V_{\mathrm{Li}}^{-}$.

\subsection{Boron-related defects}

\begin{figure}
\begin{center}
\includegraphics[width=3.2in]{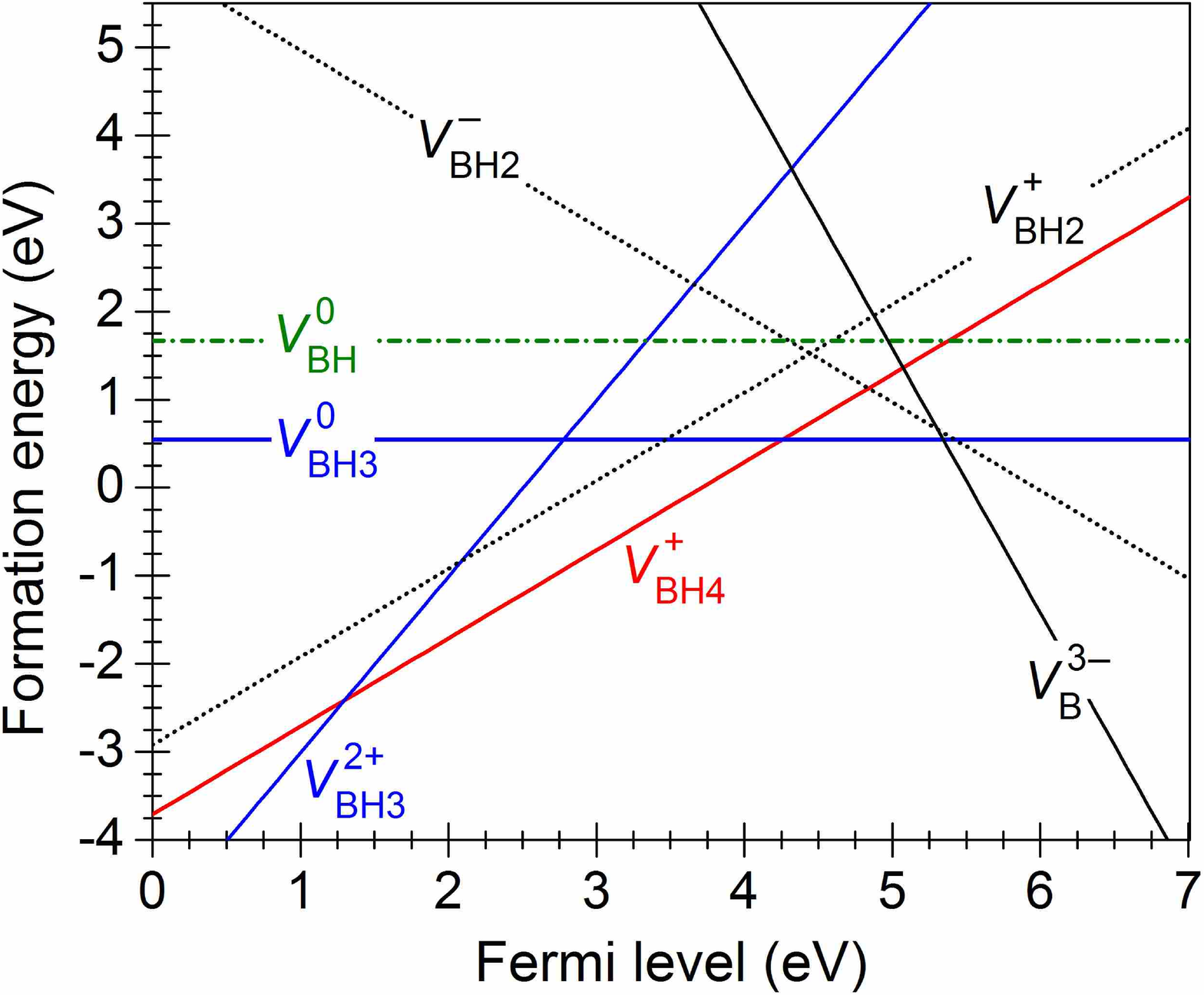}
\end{center}
\vspace{-0.2in}
\caption{(Color online) Calculated formation energies of boron-related defects in LiBH$_{4}$, plotted as a function of Fermi energy with respect to the valence-band maximum.}\label{LiBH4;FE;B}
\end{figure}

\begin{figure}
\begin{center}
\includegraphics[width=3.0in]{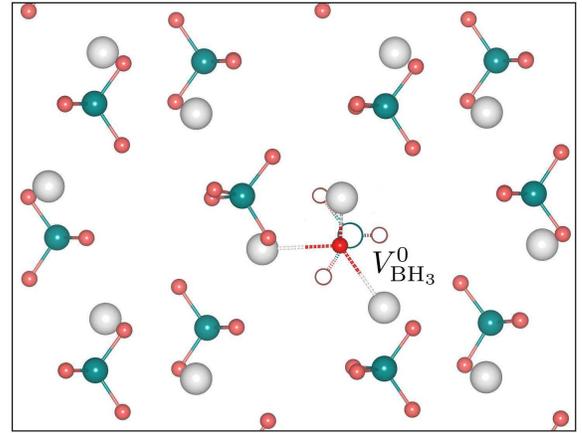}
\end{center}
\vspace{-0.15in}
\caption{(Color online) Structure of $V_{\mathrm{BH}_{3}}^{0}$ in LiBH$_{4}$. The defect can be regarded as a complex of $V_{\mathrm{BH_{4}}}^{+}$ (presented by empty spheres) and H$_{i}^{-}$ (staying near three Li$^{+}$ units).}\label{struct;VBH3}
\end{figure}

Figure \ref{LiBH4;FE;B} shows the calculated formation energies of boron vacancies ($V_{\mathrm{B}}$), BH vacancies ($V_{\mathrm{BH}}$), BH$_{2}$ vacancies ($V_{\mathrm{BH_{2}}}$), BH$_{3}$ vacancies ($V_{\mathrm{BH_{3}}}$), and BH$_{4}$ vacancies ($V_{\mathrm{BH_{4}}}$) in LiBH$_{4}$. Only stable and low-energy defects are presented. We find that $V_{\mathrm{BH_{3}}}^{2+}$, $V_{\mathrm{BH_{4}}}^{+}$, $V_{\mathrm{BH_{3}}}^{0}$, and $V_{\mathrm{B}}^{3-}$ have the lowest formation energies for certain ranges of Fermi-level values. $V_{\mathrm{BH_{4}}}^{+}$ corresponds to the removal of an entire (BH$_{4}$)$^{-}$ unit from the system. There are very small changes in the local lattice structure surrounding this defect. $V_{\mathrm{BH_{3}}}^{0}$, on the other hand, involves removing one B and three H from a (BH$_{4}$)$^{-}$ unit in LiBH$_{4}$. This results in a void formed the the removed atoms and a H$^{-}$ staying near three Li$^{+}$ units with the average Li-H distance being 1.92 {\AA}; see Fig.~\ref{struct;VBH3}. $V_{\mathrm{BH_{3}}}^{0}$, therefore, can be regarded as a complex of $V_{\mathrm{BH_{4}}}^{+}$ and H$_{i}^{-}$, with a binding energy of 1.31 eV. With such a high binding energy, even higher than the formation energy of H$_{i}^{-}$ (1.24 eV at $\mu_{e}$=4.32 eV), $V_{\mathrm{BH_{3}}}^{0}$ can occur with a concentration larger than either of its constituents under thermal equilibrium \cite{walle:3851}. The defect is, however, expected to dissociate into $V_{\mathrm{BH_{4}}}^{+}$ and H$_{i}^{-}$ at high temperatures. Like $V_{\mathrm{BH_{3}}}^{0}$, other boron-related defects can be regarded as complexes of $V_{\mathrm{BH_{4}}}^{+}$ and hydrogen-related defects. For example, $V_{\mathrm{BH}}^{0}$ is a complex of $V_{\mathrm{BH_{4}}}^{+}$, H$_{i}^{-}$, and (H$_{2}$)$_{i}$; $V_{\mathrm{BH_{2}}}^{+}$ a complex of $V_{\mathrm{BH_{4}}}^{+}$ and (H$_{2}$)$_{i}$; $V_{\mathrm{BH_{2}}}^{-}$ a complex of $V_{\mathrm{BH_{4}}}^{+}$ and 3H$_{i}^{-}$; $V_{\mathrm{BH_{3}}}^{2+}$ a complex of $V_{\mathrm{BH_{4}}}^{+}$ and H$_{i}^{+}$; and $V_{\mathrm{B}}^{3-}$ a complex of $V_{\mathrm{BH_{4}}}^{+}$ and 4H$_{i}^{-}$. The migration of $V_{\mathrm{BH_{4}}}^{+}$ involves an energy barrier of 0.27 eV, whereas for $V_{\mathrm{BH_{3}}}^{0}$, the barrier is at least 0.41 eV, given by the migration of H$_{i}^{-}$.

\section{\label{sec:disc}Discussion}

\begin{table}
\caption{Calculated formation energies ($E^{f}$) and migration energies ($E_{m}$) for selected defects in LiBH$_{4}$. Atomic chemical potentials are chosen to reflect equilibrium with LiH and H$_{2}$ gas at (1) 0 K and 0 bar and (2) 610 K and 1 bar. The formation energies for charged defects are taken at the Fermi-level position where Li$_{i}^{+}$ and $V_{\rm{Li}}^{-}$ have equal formation energies.}\label{tab:libh}
\begin{center}
\begin{ruledtabular}
\begin{tabular}{lcclc}
Defect&\multicolumn{2}{c}{$E^{f}$ (eV)}& $E_m$ (eV)&Constituents\\
&(1)&(2)&& \\
\colrule
H$_{i}^{+}$ &1.99&2.78&0.65	\\
H$_{i}^{-}$ &1.24&1.24&0.41	\\
$V_{\rm{H}}^{+}$ &1.77&1.77&0.91 \\
$V_{\rm{H}}^{-}$ &2.18&1.40&1.32 \\
(H$_{2}$)$_{i}$&0.40&1.19& -	\\
$V_{\rm 2H}$&1.11&0.33&-&$V_{\rm H}^{+}$+$V_{\rm H}^{-}$\\
Li$_{i}^{+}$	&0.50&0.50&0.30	\\
$V_{\rm{Li}}^{-}$&0.50&0.50&0.29	\\
$V_{\rm{BH_{4}}}^{+}$&0.61&0.61&0.27	\\
$V_{\rm{BH_{3}}}^{0}$&0.54&0.54&0.41\textsuperscript{\emph{a}}&$V_{\rm{BH}_{4}}^{+}$+H$_{i}^{-}$	\\
\end{tabular}
\end{ruledtabular}
\end{center}
\begin{flushleft}
\textsuperscript{\emph{a}}Lower bound, estimated by considering the defect as a complex and taking the highest of the barriers of the constituents.
\end{flushleft}
\end{table}

It emerges from our analyses in Sec.~\ref{sec:defects} that the structure and energetics of all possible native defects in LiBH$_{4}$ can be interpreted in terms of H$_{i}^{+}$, H$_{i}^{-}$, $V_{\rm{H}}^{+}$, $V_{\rm{H}}^{-}$, (H$_{2}$)$_{i}$, Li$_{i}^{+}$, $V_{\rm{Li}}^{-}$, and $V_{\rm{BH_{4}}}^{+}$, which can be regarded as elementary native defects. Table~\ref{tab:libh} summarizes key information for the most relevant native defects in LiBH$_{4}$. Defect formation energies are obtained with respect to two sets of atomic chemical potentials, assuming equilibrium with LiH and H$_{2}$ gas at (1) 0 K and 0 bar and (2) 610 K and 1 bar. Condition (1) is given here only for comparison since it does not reflect the actual experimental condition, as discussed in Sec.~\ref{sec:metho}. We find that Li$_{i}^{+}$ and $V_{\rm{Li}}^{-}$ are the charged native point defects with the lowest formation energies in both conditions. Therefore, in the absence of electrically active impurities or when such impurities occur in much lower concentrations than charged native defects, the Fermi-level position of LiBH$_{4}$ is determined by these two defects, which is at $\mu_{e}$=3.93 eV under condition (1) or 4.32 eV under condition (2) (hereafter this level will be referred to as $\mu_{e}^{\rm int}$, the Fermi-level position determined by intrinsic/native defects). We also find that the calculated formation energies of the defects, except H$_{i}^{+}$, $V_{\rm{H}}^{-}$, and (H$_{2}$)$_{i}$, are all independent of the hydrogen partial pressure and temperature, {\it cf.}~Table~\ref{tab:libh}. Since Li$_{i}^{+}$ and $V_{\rm{Li}}^{-}$ have comparable migration barriers (0.30 and 0.29 eV), we expect that they contribute almost equally to lithium-ion conductivity. The activation energy for ionic conductivity is estimated to be 0.79 eV, the summation of the formation energy and migration barrier of $V_{\rm{Li}}^{-}$, which is in agreement with the reported experimental value (0.69 eV) \cite{matsuo:224103}. Note that our calculated migration barriers of Li$_{i}^{+}$ and $V_{\rm{Li}}^{-}$ are almost equal to that (0.31 eV) for Li migration obtained in first-principles molecular dynamics simulations by Ikeshoji {\it et al.}~\cite{ikeshoji2011}.

\subsection{Decomposition mechanism}

Let us now discuss the role of native defects in the decomposition of LiBH$_{4}$ into LiH, B, and H$_{2}$ [i.e., Eq.~(\ref{eq;decomp})]. It is important to note that the decomposition necessarily involves hydrogen and/or boron mass transport in the bulk of LiBH$_{4}$. In addition, local and global charge neutrality must be maintained while charged defects are migrating. Keeping these constraints in mind, we identify the following native defects as essential to the decomposition process:

First, H$_{i}^{-}$, which is expected to act as the nucleation site for the formation of LiH as discussed in Sec.~\ref{sec:defects}. The activation energy for self-diffusion of H$_{i}^{-}$ is 1.65 eV, the sum of its formation energy and migration barrier. Note that, given the relatively high formation energy of the (H$_{i}^{-}$,$V_{\rm H}^{+}$) pair, H$_{i}^{-}$ is not likely to be created inside the material via the Frenkel pair mechanism. Other hydrogen-related defects such as H$_{i}^{+}$, $V_{\rm{H}}^{+}$, and $V_{\rm{H}}^{-}$ have very high formation energies ({\it cf.}~Table~\ref{tab:libh}) and hence low concentrations, and are expected not to play an important role. (H$_{2}$)$_{i}$ can occur with a high concentration in the bulk but it does not form by itself; below we comment on its formation in conjunction with $V_{\rm 2H}$, as suggested by Hao and Sholl \cite{hao2009}.

Second, $V_{\mathrm{BH_{4}}}^{+}$ and $V_{\mathrm{BH_{3}}}^{0}$, which are needed for boron mass transport. These two defects have the lowest formation energies in the range of Fermi-level values near $\mu_{e}^{\rm int}$, {\it cf.}~Fig.~\ref{LiBH4;FE;B}. The activation energies for the formation and migration of $V_{\mathrm{BH_{4}}}^{+}$ and $V_{\mathrm{BH_{3}}}^{0}$ are, respectively, 0.88 eV and 0.95 eV. Note that boron-related defects such as $V_{\mathrm{BH_{4}}}^{+}$ and $V_{\mathrm{BH_{3}}}^{0}$ can only be created at the surface or interface since the creation of such defects inside the material requires creation of the corresponding boron-related interstitials which are too high in energy.

Third, Li$_{i}^{+}$ and $V_{\mathrm{Li}}^{-}$, which can be created in the interior of the material in the form of a (Li$_{i}^{+}$,$V_{\mathrm{Li}}^{-}$) Frenkel pair. These low-energy and mobile point defects can act as accompanying defects in hydrogen/boron mass transport, providing local charge neutrality as H$_{i}^{-}$ and $V_{\mathrm{BH_{4}}}^{+}$ migrating in the bulk. Li$_{i}^{+}$ can also participate in mass transport that assists the formation of LiH. The activation energy for the formation and diffusion of (Li$_{i}^{+}$,$V_{\mathrm{Li}}^{-}$) in the bulk is 1.24 eV, which is the formation energy of the Frenkel pair plus the migration barrier of $V_{\mathrm{Li}}^{-}$.

Given these native defects and their properties, the decomposition of LiBH$_{4}$ can be described in terms of the following mechanism: $V_{\mathrm{BH_{4}}}^{+}$ and/or $V_{\mathrm{BH_{3}}}^{0}$ are created at the surface or interface. The creation of $V_{\mathrm{BH_{4}}}^{+}$ corresponds to removing one (BH$_{4}$)$^{-}$ unit from the bulk. Since (BH$_{4}$)$^{-}$ is not stable outside the material, it dissociates into BH$_{3}$ and H$^{-}$ where the latter stays near the surface or interface. The formation of $V_{\mathrm{BH_{3}}}^{0}$ at the surface or interface also leaves H$^{-}$ in the material and releases BH$_{3}$. It is well known that BH$_{3}$ can decompose into B and H$_{2}$, or dimerize to form diborane (B$_{2}$H$_{6}$). At room temperature, diboranes decompose to produce H$_{2}$ and higher boranes; whereas at higher temperatures (about 300$^\circ$C), diboranes decompose into B and H$_{2}$. Some diboranes or higher boranes may react with LiBH$_{4}$ to form Li$_{2}$B$_{12}$H$_{12}$ and Li$_{2}$B$_{10}$H$_{10}$ as detected in experiments \cite{cm100536a,orimo:021920,jp710894t}.

From the surface or interface, H$^{-}$ diffuses into the bulk in form of H$_{i}^{-}$ which acts as the nucleation site for the formation of LiH from LiBH$_{4}$ according to Eq.~(\ref{eq;decomp}). Here, the highly mobile Li$_{i}^{+}$ will help maintain local charge neutrality in the region near H$_{i}^{-}$. In order to maintain the reaction, (BH$_{4}$)$^{-}$ and/or BH$_{3}$ has to be transported to the surface/interface, which is equivalent to $V_{\mathrm{BH_{4}}}^{+}$ and/or $V_{\mathrm{BH_{3}}}^{0}$ diffusing into the bulk. As $V_{\mathrm{BH_{4}}}^{+}$ is migrating, local charge neutrality condition is maintained by having the highly mobile $V_{\mathrm{Li}}^{-}$ in the vacancy's vicinity. Note that, although the formation energy of $V_{\mathrm{BH_{3}}}^{0}$ is slightly lower than that of $V_{\mathrm{BH_{4}}}^{+}$ ({\it cf.}~Table~\ref{tab:libh}) and thus can occur with a higher concentration, the defect is likely to dissociate into $V_{\mathrm{BH_{4}}}^{+}$ and $V_{\rm{H}}^{-}$ at high temperatures as discussed in Sec.~\ref{sec:defects}, suggesting that these constituents may diffuse independently in the bulk of LiBH$_{4}$ during the decomposition process.

In this mechanism, the hydrogen-related diffusing species involved in the decomposition process are H$^{-}$ and (BH$_{4}$)$^{-}$ and/or BH$_{3}$ units, and the possible rate-limiting step is the formation and migration of H$_{i}^{-}$, $V_{\mathrm{BH_{4}}}^{+}$, $V_{\mathrm{BH_{3}}}^{0}$, or the (Li$_{i}^{+}$,$V_{\mathrm{Li}}^{-}$) pair in the bulk. Since H$_{i}^{-}$ gives the highest activation energy, we believe that the decomposition of LiBH$_{4}$ is rate-limited by the formation and migration of this defect. The activation energy for the decomposition of LiBH$_{4}$ is therefore 1.65 eV, the activation for self-diffusion of H$_{i}^{-}$. Our calculated value is thus higher than the reported experimental values (ranging from 1.06 to 1.36 eV) for pure LiBH$_{4}$ and the ball-milled (3LiBH$_{4}$+MnCl$_{2}$) mixture \cite{pendolino2009,Choudhury2009,varinLiBH4}, but comparable to that (1.62 eV) for LiBH$_{4}$ mixed with SiO$_{2}$ reported by Z\"{u}ttel {\it et al.}~\cite{Zuttel20031}. Note that, because the decomposition process occur at the surface or interface, ball milling that enhances the specific surface area and/or shortens the diffusion paths is expected to slightly enhance the hydrogen desorption kinetics via the kinetic prefactor.

Hao and Sholl \cite{hao2009} pointed out that the concentrations of $V_{\rm 2H}$ and (H$_{2}$)$_{i}$ can be much larger than those of $V_{\rm H}^{+}$ and H$_{i}^{-}$, and concluded that the former are the dominant hydrogen-defects in LiBH$_{4}$. Indeed, these neutral defects can have very low formation energies, as seen in Table \ref{tab:libh}.
However, $V_{\rm 2H}$ and (H$_{2}$)$_{i}$ alone cannot explain the formation of LiH in Eq.~(\ref{eq;decomp}) and as observed in experiment. In the bulk of LiBH$_{4}$, mass (and charge) conservation is required, and therefore $V_{\rm 2H}$ and (H$_{2}$)$_{i}$ can only be created in the form of a $V_{\rm 2H}$-(H$_{2}$)$_{i}$ complex, similar to a Frenkel-pair mechanism for the formation of charged hydrogen and lithium vacancies and interstitials in the bulk as presented in Sec.~\ref{sec:defects}. The activation energy associated with this complex is about 1.70 eV, obtained by adding the formation energy of $V_{\rm 2H}$-(H$_{2}$)$_{i}$ and the migration barrier of (H$_{2}$)$_{i}$, with the latter taken from Ref.~\cite{hao2009}.  This value is comparable to that associated with self-diffusion of H$_{i}^{-}$.  The formation of $V_{\rm 2H}$-(H$_{2}$)$_{i}$ complexes in the bulk is thus in principle an alternative pathway for hydrogen transport, provided there are no additional barriers involved in the formation of the $V_{\rm 2H}$-(H$_{2}$)$_{i}$ pair, an issue not addressed by Hao and Sholl \cite{hao2009}.  We note, however, that the decomposition of LiBH$_4$ via the reaction of Eq.~(\ref{eq;decomp}) necessarily involves the breaking of B-H bonds, which requires the involvement of a single-atom hydrogen-related point defect (for which we propose H$_i^-$).

\subsection{Effects of metal additives}

It should be noted that the Fermi level $\mu_{e}$ can be shifted away from $\mu_{e}^{\rm int}$, e.g., as electrically active impurities are incorporated into the material \cite{peles_prb_2007,hoang_prb_2009,wilson-short_prb_2009}. If this occurs, the formation energy, hence concentration, of charged native defects will be enhanced or reduced, depending on how the Fermi level is shifted. For $\mu_{e}$ $<$ 4.26 eV ({\it cf.}~Fig.~\ref{LiBH4;FE;B}), $V_{\mathrm{BH_{4}}}^{+}$ has a lower formation energy than $V_{\mathrm{BH_{3}}}^{0}$, and is thus expected to be the dominant defect involved in boron mass transport. Lowering $\mu_{e}$ in this range will lead to an increase (decrease) in the activation energy associated with H$_{i}^{-}$ ($V_{\mathrm{BH_{4}}}^{+}$). Since the diffusion of H$_{i}^{-}$ is the rate-limiting step, this will increase the activation energy for decomposition. Increasing $\mu_{e}$, on the other hand, will decrease the activation energy. For $\mu_{e}$ $>$ 4.26 eV, $V_{\mathrm{BH_{3}}}^{0}$ is the dominant boron-related defect. Since the activation energy associated with $V_{\mathrm{BH_{3}}}^{0}$ is independent of $\mu_{e}$, lowering (increasing) $\mu_{e}$ in this range will only increase (decrease) the activation energy associated with H$_{i}^{-}$.

As reported previously \cite{hoang_prb_2009}, transition impurities such as Ti, Zr, Fe, Ni, Pd, Pt, and Zn can be electrically active in LiBH$_{4}$ and effective in shifting the Fermi level. For example, we have found that, once Ti is incorporated into LiBH$_{4}$ at a certain lattice site with a concentration higher than that of the charged native defects, the Fermi level of the system will be determined by this impurity (hereafter referred to as $\mu_{e}^{\rm ext}$, the Fermi-level position determined by extrinsic impurities), which is at 4.16 eV (if Ti is incorporated on the B site), 4.37 eV (on the Li site), or 4.81 eV (at interstitial sites) \cite{hoang_prb_2009}. In light of the mechanism proposed above, Ti is effective in lowering the decomposition activation energy, hence enhancing the kinetics, if incorporated on the Li site or at interstitial sites (where $\mu_{e}^{\rm ext}$$>$$\mu_{e}^{\rm int}$$\equiv$4.32 eV), and ineffective if incorporated on the B site (where $\mu_{e}^{\rm ext}$$<$$\mu_{e}^{\rm int}$). Zr is expected to also enhance the kinetics of LiBH$_{4}$ because we find that the incorporation of Zr on the Li and B sites and at the interstitial sites give rise to $\mu_{e}^{\rm ext}$=4.60, 4.44, and 4.45 eV \cite{hoang_prb_2009}, which are all higher than $\mu_{e}^{\rm int}$. On the other hand, we find that Fe, Ni, Pd, Pt, and Zn give rise to $\mu_{e}^{\rm ext}$$<$$\mu_{e}^{\rm int}$ \cite{hoang_prb_2009}, and are thus expected not to be effective additives. These impurities may, however, lower the formation energy of $V_{\mathrm{BH_{4}}}^{+}$ and thus enhance the release of BH$_{3}$ and probably reduce the onset decomposition temperature.

Experimentally, it has been found that Ti- and Zn-containing additives such as TiO$_{2}$ \cite{Yu_LiBH4_2008}, TiCl$_{3}$ or TiF$_{3}$ \cite{Au2008,fang_2011}, and ZnF$_{2}$ are effective in enhancing the kinetics \cite{Au2008}, whereas FeCl$_{3}$ is ineffective \cite{Au2008}. Our results are therefore in agreement with the available experimental data in the case of Ti and Fe, whereas there is a disagreement in the case of Zn. One possible explanation for this discrepancy is that we have not yet investigated all possible interactions between Zn and the constituents of the host material. The halogen anion in ZnF$_{2}$ may also play a role, as it was reported to be the case for titanium halides \cite{fang_2011}. Finally, the addition of Ni to nanoconfined LiBH$_{4}$ has been found to slightly lower the onset temperature of hydrogen release but does not have a significant effect on the hydrogen desorption \cite{ngene_2011}, which appears to be consistent with our conclusions for Ni discussed above. Note that it was reported more recently that mixing or ball milling of LiBH$_{4}$ with FeCl$_{2}$ resulted in a significant decrease in the decomposition temperature \cite{zhang2010}, which seems to contradict to the previous report of LiBH$_{4}$ mixing with FeCl$_{3}$ \cite{Au2008}. Further theoretical and experimental studies are therefore needed to clarify this situation.

It is important to note that the incorporation of metal impurities and the formation of native point defects are processes that are very distinct and occur in different stages of preparation or use of the material. The point defects are formed during decomposition, in a process that is close to equilibrium, such that their concentration will be determined by their formation energy $-$ which is quite low, as seen in Table \ref{tab:libh}. The metal impurities, on the other hand, are incorporated during initial processing of the material, often in a process such as ball milling, which can be highly energetic and potentially introduce impurities in non-equilibrium concentrations not directly related to their formation energy. This allows for incorporation of impurities with formation energies higher than those of the point defects.

\section{\label{sec:sum}Summary}

We have carried out a comprehensive first-principles study of native point defects and defect complexes in LiBH$_{4}$. We find that lithium vacancies and interstitials have low formation energies and are highly mobile. These defects can participate in lithium-ion conduction, and act as accompanying defects in hydrogen and boron mass transport. We have proposed a specific mechanism for the decomposition of LiBH$_{4}$ that involves the formation and migration of H$_{i}^{-}$, $V_{\mathrm{BH_{4}}}^{+}$, $V_{\mathrm{BH_{3}}}^{0}$, and (Li$_{i}^{+}$,$V_{\mathrm{Li}}^{-}$) in the bulk LiBH$_{4}$. Based on this atomistic mechanism, we explain the decomposition and dehydrogenation, the rate-limiting step in hydrogen desorption kinetics, and the effects of metal additives on the kinetics of LiBH$_{4}$.

\begin{acknowledgments}

This work was supported by General Motors Corporation and by the U.S.~Department of Energy (Grant No. DE-FG02-07ER46434). It made use of NERSC resources supported by the DOE Office of Science under Contract No.~DE-AC02-05CH11231 and of the CNSI Computing Facility under NSF Grant No.~CHE-0321368. The writing of this paper was partly supported by the Naval Research Laboratory through Grant No.~NRL-N00173-08-G001.

\end{acknowledgments}


\end{document}